\documentclass[twocolumn,,aps,prb,showpacs,amsmath,superscriptaddress]{revtex4-1}
\usepackage{}
\usepackage{txfonts}
\usepackage{color}
\usepackage{amssymb}
\usepackage{mathrsfs}
\usepackage{graphicx,epstopdf}
\usepackage{float}

\begin{document}


\begin{abstract}
One dimensional Cherenkov processes in ferromagnetic isolators are studied with perturbation theory under the constraint condition of conservation of energy and momentum. It is shown that the magnon-phonon interaction channels are limited and wave number dependent, which result in respectively $1/k^{2}$ and $1/k^{4}$ dependence of the lifetime and the relaxation time of long wavelength magnons. The reciprocal of relaxation time between magnons and phonons, $1/\tau_{mp}$, is found to be a linearly increasing function of the temperature  as $T>$ 70 K. Based on the Sanders-Walton model,  we further show that when a thermal (phonon) gradient is applied along the system, the temperature difference between the phonon bath and the magnons with wave-vector $k$ becomes more pronounced as $k$ decreasing.
\end{abstract}

\title{One dimensional Cherenkov processes in ferromagnetic insulator}
\author{Tongli Wei}
\email{weitl12@lzu.edu.cn}
\affiliation{Key Laboratory for Magnetism and Magnetic Materials of Ministry of Education, Lanzhou University, Lanzhou 730000, China.}
\affiliation{Key Laboratory of Physics and Photoelectric Information Functional Materials Sciences and Technology, College of Electric and Information Engineering, North Minzu University, Yinchuan 750021, China.
}

\author{Yaojin Li}

\affiliation{Key Laboratory for Magnetism and Magnetic Materials of Ministry of Education, Lanzhou University, Lanzhou 730000, China.}
\author{Decheng Ma}

\affiliation{Key Laboratory for Magnetism and Magnetic Materials of Ministry of Education, Lanzhou University, Lanzhou 730000, China.}
\author{Chenglong Jia}
\email{cljia@lzu.edu.cn}
\affiliation{Key Laboratory for Magnetism and Magnetic Materials of Ministry of Education, Lanzhou University, Lanzhou 730000, China.}

\date{}
\maketitle

\section{Introduction}
Exerting a temperature gradient to generate a spin voltage, known as the spin Seebeck effect is becoming a very promising direction in spintronics \cite{Bauer2012, Maekawa2016, Maekawa2008, Maekawa2010, Jaworski2010, Uchida2011}. It has been revealed by the experimental observations of the long length scale ($\sim$ several millimeters) spin current in ferromagnetic metals \cite{Maekawa2008} and the spin Seebeck effect in magnetic insulators  \cite{Maekawa2010, Uchida2011}  that the dynamics of localized spins in the ferromagnets, i.e., magnons rather than the conduction electrons are essential carriers for the spin Seebeck effect.  The non-equilibrium dynamics of magnons and phonons (as well as their couplings) are believed to play a key role in the spin Seebeck effect \cite{XiaoJiang2010, Adachi2011, Uchida2012, Adachi2013, Maekawa2016, Tatara2010, Bauer2013, Hoffman2013, Ritzmann2014}. To start with the assumption that the magnon and the phonon can establish dynamic thermal equilibria at distinct temperatures, $T_m$ and $T_p$, respectively,  a linear-response theory of the spin Seebeck effect was developed by Maekawa's group and other researchers \cite{ Maekawa2010, Uchida2011, XiaoJiang2010, Maekawa2016, Adachi2011,  Adachi2013}. These theoretical models have successfully described the general properties of the spin Seebeck effect. The characteristic length $\lambda$, which is proportional to the square root of the magnon-phonon relaxation time, is theoretically estimated to be in the rang of $ 0.85-8.5$ mm \cite{XiaoJiang2010}. However, the direct spatially resolved measurement of the magnon temperature in yttrium iron garnet (YIG) by M. Agrawal \emph{et al}\cite{Agrawal2013} via the Brillouin light scattering
(BLS) spectroscopy indicates that there is no considerable temperature difference  between phonons and the short wavelength thermal magnons ($\geq 10^{8}$rad/m). The maximum value of $\lambda$ deduced from the difference between $T_m$ and $T_p$ is $0.47$mm. To resolve this puzzle, they argued that the long wavelength magnons  ($< 10^{7}$rad/m) may have large temperature difference with the phonon bath and are primarily responsible for the spin Seebeck effect. Recently, several experiments with the setup of YIG/Pt thin film bilayers have been launched to clarify this problem. It is shown that the cutoff frequency of microwaves to generate spin injection by spin Seebeck effect  is up to 30MHz \cite{Roschewsky2014} or 6.875GHz \cite{Schreier2016}, depending on the thickness of the YIG film. The mean free path, the characteristic frequency and the relaxation time have been  analyzed  theoretically as well \cite{Heremans2014, Etesami2015, Diniz2016}. It is agreed that spin Seebeck effect is spectral dependent. The relaxation time and temperature distribution of long-wavelengh magnons may be different from the ones of short wavelength magnons \cite{Diniz2016, Ruckriegel2014, Kikkawa2015, Heremans2015, Kikkawa2013, Kehlberger2015, ManHaoran2017}.

In this paper the magnon-phonon interaction conditioned by the exchange-striction effect is detailly analyzed and studied. In section \uppercase\expandafter{\romannumeral2} we revealed that the strength of magnon-phonon coupling is spectral dependent. The transition and relaxation processes are investigated in section \uppercase\expandafter{\romannumeral3} and \uppercase\expandafter{\romannumeral4}. It is found that, under the constraint condition of conservation of energy and momentum, the transition probabilities, the relaxation times and the transition channels display a strong wave number dependence. In section \uppercase\expandafter{\romannumeral5}, temperature distributions of  
magnons is solved with the Sanders-Walton (SW) model \cite{Sanders1977}. It is evident that, along the direction of thermal gradient, the temperature distribution of magnons is largely lifted up in the cooler regions and deeply pulled down in the hotter ones as the wave number of magnon decreasing. It suggests that the longer wavelength magnons can have larger temperature deviations from the phonons due to the relatively weak coupling between them. The characteristic magnon wave vector in YIG is estimated  as $1.68\times 10^{8}$ rad/m.

\section{Magnon-phonon coupling}
We provide a quantitative justification of coupling between magnons and phonons by analyzing the Heisenberg Hamiltonian
\cite{L.Z.Zhong2001, C. Kittel, Huangkun1988,Adachi2013,Jia2009,Ruckriegel2014}:
\begin{equation}
\mathcal{H}= - \sum_{\langle ij \rangle}J_{ij}\hat{S}_{i}\cdot \hat{S}_{j},
\end{equation}
where  $\hat{S}_{i}$ is the spin operator at site $i$ and the sum is  made over the nearest neighbors. $J_{ij}> 0$ describes the ferromagnetic exchange interaction that depends on the distance for a pair of spins, i.e., $J_{ij}= J(|\textbf{R}_{i}-\textbf{R}_{j}|) $ is not a constant but changes with the vibration of lattice, which is called the exchange-striction effect.

By using  Hostein-Primakoff transformation \cite{L.Z.Zhong2001, C. Kittel, Huangkun1988}, ignoring the high{er}-order terms,  we arrive at the Hamiltonian of spin wave (magnons),
\begin{eqnarray}
H_{m}&=&\sum_{\textbf{k}}\hbar \omega^{(m)}_{\textbf{k}}b^{\dag}_{\textbf{k}}b_{\textbf{k}},
\end{eqnarray}
with $b^{\dag}_{\textbf{k}}$ and $b_{\textbf{k}}$ being the creation and annihilation operators of a magnon with wave-vector $\textbf{k}$ (i.e., $\textbf{k}$-magnon), respectively. The dispersion relation of magnons along a ferromagnetic chain reads $\hbar \omega_{k}^{(m)}\approx 2J_{0}S(1-\cos{ka}) = \mathcal{J}(1-\cos{ka})$, where $\hbar$ is Planck's Constant, $\omega_{k}^{(m)}$ is the frequency of $\textbf{k}$-vector  magnons, $J_0$ is the equilibrium exchange integral, $S$ is the average spin of the magnetic ion and $a$ is the lattice constant of the one-dimensional ferromagnetic insulator.

For vibrations of the lattice, the harmonic approximation can be used \cite{L.Z.Zhong2001, C. Kittel, Huangkun1988}, we have then the Hamiltonian of lattice displacement (phonons),
\begin{eqnarray}
H_{p}&=&\sum_{\textbf{q}}\hbar \omega^{(p)}_{\textbf{q}}p^{\dag}_{\textbf{q}}p_{\textbf{q}},
\label{eq:Hamilton-of-phonons-conserved-self-energy}
\end{eqnarray}
where $\hbar\omega^{(p)}_{\textbf{q}}$ is the energy quantum of $\textbf{q}$-vector phonons,  $p^{\dag}_{\textbf{q}}$/$p_{\textbf{q}}$ is the creation/annihilation operators of $\textbf{q}$-phonon. Considering the longitudinal acoustic branches, the phonon dispersion relation is given by $\hbar\omega^{(p)}_{\textbf{q}}=\mathcal{D}\sin{\frac{|qa|}{2}}$ \cite{L.Z.Zhong2001, C. Kittel, Huangkun1988}.

Supposing the exchange integral falls off as a power law with the separation of the magnetic ions\cite{Adachi2013,Jia2009}, we have $J_{ij}= J_{0}+\nabla J \cdot (\textbf{u}_{i}-\textbf{u}_{j})$, $\textbf{u}_{i}$ and $\textbf{u}_{j}$ are the displacements of lattice $i$ and $j$. The magnon-phonon interaction is given by \cite{Adachi2013}
\begin{eqnarray}
H_{\text{mp}}&\approx& {-}\sum_{i,j}J_{0}\frac{g}{a}\textbf{e}_{ij}\cdot (\textbf{u}_{i}-\textbf{u}_{j})\hat{S}_{i}\cdot \hat{S}_{j},
\label{eq:ragulation-of-exchange-striction-to-heisenberg-model}
\end{eqnarray}
where $\textbf{e}_{ij}$ is the unit vector of site $i$  to site  $j$, the factor $g$ ($6-14$ Ref.[\onlinecite{Harrison1980}]) is defined as the relative rate-of-change of $J_{ij}$ to the difference of the displacement of lattice $\textbf{u}_{i}$ and $\textbf{u}_{j}$\cite{Adachi2013,Jia2009},
$\nabla (J/J_{0}) =(g/a)\textbf{e}_{ij}$. $\textbf{u}_{l}$  in the $\textbf{q}$-space reads, \cite{Adachi2013}
\begin{eqnarray}
\textbf{u}_{l}&=& \text{i}\sum_{\textbf{q}}\sqrt{\frac{\hbar}{2 \omega^{(p)}_{\textbf{q}} NM_{ion}}}
\textbf{e}_{\textbf{q}}\mathcal{B}_{\textbf{q}}e^{\text{i}\textbf{q}\cdot \textbf{R}_{l}},
\label{eq:displacement-of-lattice}
\end{eqnarray}
where $N$ is the number of the sites, $M_{ion}$ is the mass of the magnetic ions, and $\mathcal{B}_{\textbf{q}}=p_{\textbf{q}}+p^{\dag}_{-\textbf{q}}$, $\textbf{R}_{l}$ is the position vector of lattice $l$. Taking $U_{l,\delta}=\textbf{e}_{\delta}\cdot(\textbf{u}_{l}-\textbf{u}_{l+\delta})$ to the first order of Taylor's expansion, where $l+\delta$ denotes the nearest site of site $l$, we have
\begin{eqnarray}
U_{l,\delta}&=& \sum_{\textbf{q}}\sqrt{\frac{\hbar}{2 \omega^{(p)}_{\textbf{q}}NM_{ion}}}
\mathcal{B}_{\textbf{q}}e^{i\textbf{q}\cdot \textbf{R}_{l}}(\textbf{e}_{\textbf{q}}\cdot \textbf{e}_{\delta})(\textbf{q}\cdot a\textbf{e}_{\delta}),
\end{eqnarray}
transverse models are excluded here. Given $\sum_{l,\delta}S^{2}U_{l,\delta}=0$, the magnon-phonon interaction Eq.(\ref{eq:ragulation-of-exchange-striction-to-heisenberg-model}) can be rewritten as
\begin{eqnarray}
H_{\text{mp}}
&&= \sum_{l,\delta}J_{0}\frac{gS}{a}U_{l,\delta}
(b^{\dag}_{l}b_{l}+b^{\dag}_{l+\delta}b_{l+\delta}-b_{l}b^{\dag}_{l+\delta}-b^{\dag}_{l}b_{l+\delta})\nonumber\\
&&= \sum_{k,q} V_{mp}(k,q)\mathcal{B}_{q}b^{\dag}_{k+q}b_{k},
\label{eq:couplings-of-phonon-magnon-system}
\end{eqnarray}
here $b^{\dag}_{l}$/$b_{l}$ is the creation/annihilation operator of spin deviation at site $l$. The strength of the coupling is expressed as
\begin{eqnarray}
V_{mp}(k,q)  &=& \frac{ g \mathcal{J}}{\sqrt{N}}\sqrt{\frac{\hbar \omega^{(p)}_{q}}{2 M_{ion}v^{2}_{p}}} \nonumber \\ &&[1-\cos{ka}+\cos{qa}-\cos{(k+q)a}],
\end{eqnarray}
where $v_{p}$ is the velocity of the phonons. Note that here only longitude phonons are coupled to the magnons. More importantly, it is clear the coupling strength $V_{mp}(k,q)$ is wave number dependent and the coupling of long-wavelength magnons can be sufficiently weak for them to contribute to the spin Seebeck effect.

\section{Transition probabilities induced by Cherenkov processes}
\begin{figure}[b]\centering
\includegraphics[scale=0.27]{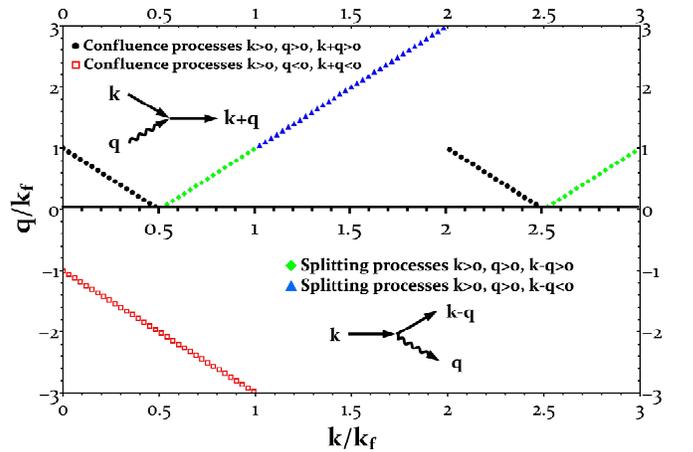}
\caption{\textbf{{Cherenkov channels of $k$-magnons in one dimensional  processes. }}\footnotesize{The conservation of energy and momentum has been used to constrain the transition channels, involving a $q$-phonon and two magnons with the wave-vector $k$ and $k\pm q$, respectively.
$k_{f}=\frac{2}{a}\arcsin{\frac{\mathcal{D}}{2\mathcal{J}}}$ is the cross-over point of the magnon's and phonon's dispersion relations, i.e., $\omega^{(m)}_{k_f} = \omega^{(p)}_{k_f}$.  Here $\mathcal{D}=\mathcal{J}$ is used.}}
\label{fig:Conservation-law-of-three-particle-process}
\end{figure}
Clearly, the exchange-striction effect in ferromagnetic materials only gives rise to the three-particle Cherenkov processes \cite{Gurevich1996}, i.e., the elementary processes of splitting a magnon into magnon and phonon, or the reverse ones (the {magnon-phonon} confluence with creation of a magnon), there is no processes of two-magnon confluence into a phonon and/or the splitting a phonon into two magnons.

While the Cherenkov processes does not change the number of total magnons, it can result in a redistribution of magnons in the $k$-space. Considering that the energy is conserved{,\cite{ Kaganov1959,Gurevich1996,Ruckriegel2014}}
\begin{eqnarray}
\omega^{(p)}_{q}+\omega^{(m)}_{k}= \omega^{(m)}_{k+q} \hspace{0.5cm}\text{or}\hspace{0.5cm} \omega^{(m)}_{k} = \omega^{(m)}_{k-q}+\omega^{(p)}_{q}
\label{eq:condition-of-conservation-of-energy-2}
\end{eqnarray}
for the confluence or splitting processes, respectively. By substitute the dispersion relaxations to Eq.(\ref{eq:condition-of-conservation-of-energy-2}), we have then
\begin{eqnarray}
&&\cos{(k+q)a}=(\cos{ka}-\frac{\mathcal{D}}{\mathcal{J}}\sin{|qa/2|})\hspace{0.5cm}\text{or}\nonumber\\
&&\cos{(k-q)a}=(\cos{ka}+\frac{\mathcal{D}}{\mathcal{J}}\sin{|qa/2|}).
\label{eq:Conservation-law-of-three-particle-process}
\end{eqnarray}
The Cherenkov processes can happen just in some particular wave-vector channels, as shown in Fig.\ref{fig:Conservation-law-of-three-particle-process}. In particular, we note that some magnons are only directly involved in one of the Cherenkov processes, which would influence the energy transfer efficiency between the magnons and phonons.

The transition probabilities can be calculated within the perturbation approximation. The probability of transiting from state  $|\alpha \rangle$ to state $|\alpha'\rangle$ (with equal or approximately equal energy) in unit time  can be given by \cite{Dirac1958, Landau3, Kaganov1958, Kaganov1959, Ziman1960, Gurevich1996}
\begin{eqnarray}
p(\alpha\alpha')=\frac{2\pi}{\hbar }|\langle \alpha'|\mathcal{V}|\alpha \rangle|^{2}\mathcal{O}(\varepsilon_{\alpha}-\varepsilon_{\alpha'}),
\label{eq:}
\end{eqnarray}
where $\mathcal{V}$ is the perturbation operator,  $\varepsilon_{\alpha}-\varepsilon_{\alpha'}$ is the energy difference between the states of $|\alpha \rangle$ and $|\alpha' \rangle$. $\mathcal{O}(\varepsilon)\equiv\frac{\sin{\varepsilon t /\hbar}}{\pi \varepsilon}$, we take it as a constant to all Cherenkov processes with $\mathcal{O}(\varepsilon)=1/\Delta_{3}$ in the present study. For the processes of a $k$-magnon and a $q$-phonon confluence  into a $(k+q)$-magnon, the initial state can be written as $|\alpha\rangle=|\cdots n^{(m)}_{k}n^{(m)}_{k+q}n^{(p)}_{q} \cdots \rangle$, and the final state is $|\alpha'\rangle=|\cdots (n^{(m)}_{k}-1)(n^{(m)}_{k+q}+1)(n^{(p)}_{q}-1) \cdots \rangle$, where $n^{(m)}_{k},n^{(m)}_{k+q},n^{(p)}_{q}$ {are} Bose-Einstein distributions of $k, ~k+q$ magnons and $q$-phonon {\cite{Kaganov1959}}. The transition probabilities of the confluence processes reads then
\begin{eqnarray}
p^{(c)}_{3}(k,q)&=&\frac{2\pi}{\hbar \Delta_{3}}|\langle \alpha'|V_{mp}(k,q)p_{q}b^{\dag}_{k+q}b_{k}|\alpha \rangle|^{2}\nonumber\\
&=&\frac{2\pi g^{2}}{N\hbar \Delta_{3}}\frac{\mathcal{J}^{2}[1-\cos{ka}+\cos{qa}-\cos{(k+q)a}]^{2}}{2 M_{ion}v^{2}_{p}}\nonumber\\
&&\hspace{1cm}n^{(m)}_{k}(n^{(m)}_{k+q}+1)n^{(p)}_{q}\hbar \omega^{(p)}_{q}.
\label{eq:jumping-probabilities-of-three-particle-1}
\end{eqnarray}
Whereas, for the processes of a $k$-magnon splitting to a $(k-q)$-magnon and a $q$-phonon, the final state is $|\alpha''\rangle=|\cdots (n^{(m)}_{k}-1)(n^{(m)}_{k-q}+1)(n^{(p)}_{q}+1) \cdots\rangle$. The transition probabilities is
\begin{eqnarray}
p^{(s)}_{3}(k,q)&=&\frac{2\pi g^{2}}{N\hbar \Delta_{3}}\frac{\mathcal{J}^{2}[1-\cos{ka}+\cos{qa}-\cos{(k-q)a}]^{2}}{2 M_{ion}v^{2}_{p}}\nonumber\\
&&\hspace{1cm}n^{(m)}_{k}(n^{(m)}_{k-q}+1)(n^{(p)}_{q}+1)\hbar \omega^{(p)}_{q}.
\label{eq:jumping-probabilities-of-three-particle-2}
\end{eqnarray}
The averaged transition probability of annihilating a $k$-magnon through the Cherenkov processes at temperature $T$ is thus given by
\begin{eqnarray}
P_{3}(k,T)&=&\frac{\sum_{q}p^{(c)}_{3}(k,q)+\sum_{q'}p^{(s)}_{3}(k,q')}{n^{(m)}_{k}}.
\label{eq:relative-probabilities-of-magnons}
\end{eqnarray}

\begin{figure}[t]\centering
\includegraphics[scale=0.26]{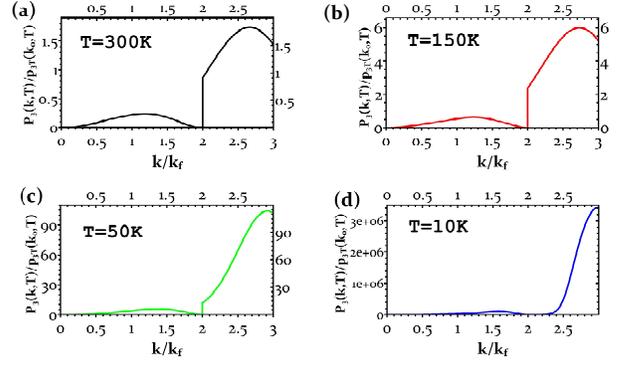}
\caption{\textbf{Relative transition probabilities of annihilating a $k$-magnon}. \footnotesize{Here $\mathcal{D}=\mathcal{J}=19$mev.}}
\label{fig:lifetime-and-jumping-probabilities-for-long-wave-vector-magnons}
\end{figure}

At room temperatures, Boltzmann's theorem of the equipartition of energy can be used and the heat capacity of every wave vector can be considered as $k_{B}${\cite{Huangkun1988,Ruckriegel2014}}, where $k_{B}$ is the Boltzmann's Constant. We take the rough approximations as follows:
$(n^{(p)}_{q}+1)\hbar \omega^{(p)}_{q}\sim k_{B}T$, $(n^{(m)}_{k+q}+1)\hbar \omega^{(m)}_{k+q}\sim k_{B}T$, and $n^{(m)}_{k}\hbar \omega^{(m)}_{k}\sim k_{B}T$. On the other hand, as demonstrated in Fig.\ref{fig:Conservation-law-of-three-particle-process}, the long wavelength magnons are just coupled with phonons with the wave vector very close to $k_{f}$ or $-k_{f}$, where $k_{f}$ is the cross-over point of their dispersion relations, satisfying $k_{f}=\frac{2}{a}\arcsin{\frac{\mathcal{D}}{2\mathcal{J}}}$, and therefore, one has: $\cos{(k\pm q)a}\approx \cos{qa}\mp ka\sin{qa}$. The total probability due to the Cherenkov confluence processes of creating a $\pm k_f$ magnon with annihilating a long wavelength ($k/k_f \ll 1$) magnon can be approximated as,
\begin{eqnarray}
p_{3T}(k,T)&=&p^{(c)}_{3}(k,q_{1})+p^{(c)}_{3}(k,q_{2})\nonumber\\
&\approx&\frac{8\pi g^{2}}{N\hbar \Delta_{3}}\frac{1+\cos{k_{f}a}}{2 M_{ion}v^{2}_{p}}(k_{B}T)^{3},
\label{eq:Total-jumping-probabilities-of-k-magnons}
\end{eqnarray}
where $q_{1}, q_{2}$ are the phonon wave vectors of the two transition channels near $k_{f}$ and $-k_{f}$. Since the total probability $p_{3T}(k)$ of annihilating processes of long wavelength magnons  is only temperature dependence, we can take it as a reference to calculate the spectral dependent transition probabilities given by the Eqs. (\ref{eq:jumping-probabilities-of-three-particle-1}), (\ref{eq:jumping-probabilities-of-three-particle-2}) and (\ref{eq:relative-probabilities-of-magnons}). The results are shown in Fig. \ref{fig:lifetime-and-jumping-probabilities-for-long-wave-vector-magnons}, which is qualitatively consistent with the damping factor  of YIG (stemmed from the Cherenkov processes as well) calculated by A. R$\ddot{\text{u}}$ckriegel \emph{et al} in Ref.[\onlinecite{Ruckriegel2014}]: two different regimes with distinct peaks (around  $1.3k_{f}$ and $2.7k_{f}$, respectively).

Treat magnon as a quasi-particle, the lifetime of long wavelength magnons $(|k|\ll k_{f})$ determined by the Cherenkov processes reads then \cite{Landau3}
\begin{eqnarray}
\tau'_{mk}&=&\frac{1}{P_{3}(k,T)}\approx\frac{N\hbar \Delta_{3} M_{ion}v^{2}_{p}}{2\pi g^{2}\mathcal{J}(1+\cos{k_{f}a})}\frac{1}{(k_{B}T)^{2}}
\frac{1}{(ka)^{2}}.
\label{eq:lifetime-of-magnons-2}
\end{eqnarray}
The self-energy is given by the uncertainty principle,  $\Delta_{mk}=\hbar/\tau'_{mk}\propto T^{2}k^{2}$. Clearly, the averaged relaxation probability of the Cherenkov processes is $k$-dependent.

\section{Magnon/phonon relaxation}
For the Cherenkov processes considered here, energy is exchanged between the magnon and phonon subsystems. When the magnon and phonon subsystems are in the statistical equilibrium, the net energy transport is zero. However, when two subsystems aren't in statistical equilibrium, relaxations stemmed from the imbalance of Cherenkov confluence processes and splitting processes will be triggered. M. I. Kagnov and V. M. Tsukernik have pointed out that the  equilibrium will be established first within each subsystem (magnons and phonons), and then considerably more slowly between the two subsystems \cite{Kaganov1959}. In the present study, however, given that the strength of the coupling and the transition probability are spectral dependent, the establishment of the equilibrium will be more complex, we need to examine the relaxation processes of each $k$-magnon.

\subsection{Relaxation time of long wavelength magnons}
Firstly, the magnon relaxation is studied by exploring the imbalance of the $k$-magnons in different Cherenkov processes.
Considering that a $k$-magnon, due to thermal perturbations or fluctuations, is excited to temperature $T_{mk}$, while the rest magnons remain in the equilibrium with temperature $T_{m}$. 
and the temperature difference $\Delta T_{mk}=T_{mk}-T_{m}$ satisfies $|\Delta T_{mk}|/T_{m}\ll 1$. By using the relaxation equation \cite{Sanders1977}
\begin{eqnarray}
\frac{d (T_{mk}-T_{m})}{dt}=-\frac{1}{\tau_{mk}}(T_{mk}-T_{m}),
\end{eqnarray}
we have
\begin{eqnarray}
\frac{d T_{mk}}{dt}=-\frac{1}{\tau_{mk}}(T_{mk}-T_{m}).
\label{eq:equation-of-ralaxation}
\end{eqnarray}
The relaxation time $\tau_{mk}$ is given by \cite{Sanders1977}
\begin{eqnarray}
\tau_{mk}&=&-c_{mk}\frac{T_{mk}-T_{m}}{d\mathbb{Q}_{k}/dt},
\label{eq:ralaxation-time-of-k-vector-magnons}
\end{eqnarray}
where $c_{mk}$ is the heat capacity of $k$-magnons, and $\mathbb{Q}_{k}$ describes the  energy gain
of $k$-magnons. {The rate of the energy gain $d\mathbb{Q}_{k}/dt$ can be} written as\cite{Kaganov1958, Kaganov1959}
\begin{eqnarray}
\frac{d\mathbb{Q}_{k}}{dt}=[p^{(i)}_{3T}(k,T)-p_{3T}(k,T)]\hbar \omega^{(m)}_{k},
\end{eqnarray}
$p^{(i)}_{3T}(k,T)$ is the total probability that  a $k$-magnon is created. {In the case of statistical equilibrium, the energy gain is expected to be zero \cite{Kaganov1958, Kaganov1959} and $p_{3T}(k,T)=p^{(i)}_{3T}(k,T)$ is valid. However, when the temperatures is not equal, $p_{3T}(k,T)\neq p^{(i)}_{3T}(k,T)$. To leading terms, we have $n^{(m)}_{k}(T_{mk})=n^{(m)}_{k}(T_{m})+\frac{dn^{(m)}_{k}(T_{m})}{dT}\Delta T_{mk}$, the difference of the probabilities for long wavelength magnons can be deduced as
\begin{eqnarray}
p^{(i)}_{3T}(k,T_{mk})-p_{3T}(k,T_{mk})
\approx-p_{3T}(k,T_{m})\frac{\hbar \omega^{(m)}_{k}}{k_{B}T_{m}}\frac{\Delta T_{mk}}{T_{m}}.
\end{eqnarray}
Eq.(\ref{eq:ralaxation-time-of-k-vector-magnons}) gives that
\begin{eqnarray}
\tau_{mk}&\approx&{\frac{c_{mk}}{k_{B}p_{3T}(k)}\left(\frac{k_{B}T_{m}}{\hbar \omega^{(m)}_{k}}\right)^{2}}\nonumber\\
&\approx&\frac{N\hbar \Delta_{3} M_{ion}v^{2}_{p}}{\pi g^{2}(1+\cos{k_{f}a})\mathcal{J}^2}\frac{1}{k_{B}T}\frac{1}{(ka)^{4}},
\label{eq:ralaxation-of-k-wave-vector-magnons}
\end{eqnarray}
where $c_{mk}\approx k_{B}$ has been used.

Eq.(\ref{eq:ralaxation-of-k-wave-vector-magnons}) shows that $1/\tau_{mk}$  is proportional to $k^{4}$, which reduces rapidly with reducing $k$.
For short wavelength magnons, $\tau_{mk}$ can be much smaller than the relaxation time $\tau_{mp}$ between magnons and phonons,  equilibrium will be established within the short wavelength magnons firstly. But for the magnons with condition $\tau_{mk}\gg \tau_{mp}$, they can't  catch up with the relaxation between the magnon and phonon subsystems. Consequently, the temperatures of magnons becomes $k$-dependent. By defining the wave vector  $k_{c}$ as $\tau_{mk_{c}}=\tau_{mp}$, correspondingly, relaxations of the non-equilibrium  magnon-phonon system  can be processed in three stages. In the first stage,  equilibrium will be established in short wavelength magnons (with the condition $k\gg k_{c}$).  In the second stage,  equilibrium will be established between phonons and short wavelength magnons ($k\gg k_{c}$). In the third stage, equilibrium will be established between long wavelength magnons ($k\ll k_{c}$) and the rest short wavelength magnons of the system.

\begin{figure}[b]
\includegraphics[scale=0.38]{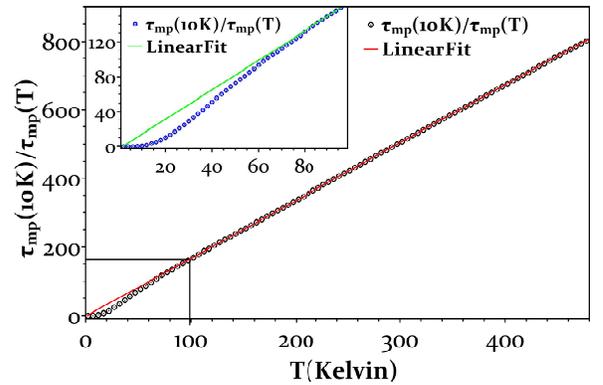}
\caption{\textbf{The dependence of relaxation time between magnons and phonons on  the temperature T.
 }\footnotesize{Here $\mathcal{J}=\mathcal{D}$, and the velocity of the phonons is set to be a constant.
}}
\label{fig:relaxation-time-for-three-particle-process}
\end{figure}

\subsection{Relaxation between magnons and phonons}
Considering  phonons with  temperature $T_{p}$ and the (short wavelength) magnons with temperature $T_{m}$, the difference is $\Delta T_{pm}=T_{p}-T_{m}$. By using the relaxation equation\cite{Sanders1977}
\begin{eqnarray}
\frac{d T_{p}}{dt}=-\frac{c_{m}}{c_{T}}\frac{1}{\tau_{mp}}(T_{p}-T_{m}),
\label{eq:equation-of-ralaxation-phonons}
\end{eqnarray}
where $c_{T}=c_{p}+c_{m}$ is the total heat capacity of the system, $c_{m}, c_{p}$ are the heat capacities of short wavelength
magnons and phonons, respectively, the reciprocal of the mean relaxation time is obtained\cite{Sanders1977}
\begin{eqnarray}
\frac{1}{\tau_{mp}}&=&-\frac{c_{T}}{c_{p}c_{m}}\frac{1}{\Delta T_{pm}}\sum_{q}\frac{d\mathbb{Q}_{q}}{dt}\\
&=&-\frac{c_{T}}{c_{p}c_{m}}\frac{1}{\Delta T_{pm}}\sum^{|k+q|>k_{c}}_{|k|>k_{c},q}[ p^{(s)}_{3}(k+q,q)- p^{(c)}_{3}(k,q)]\hbar \omega^{(p)}_{q},\nonumber
\label{eq:ralaxation-time-of-phonon-magnon-system}
\end{eqnarray}
$d\mathbb{Q}_{q}/dt$ is the energy gain of $q$-vector phonons in unit time, $p^{(s)}_{3}(k+q,q)$  ($p^{(c)}_{3}(k,q)$) is the probability of the creation (annihilation) of a $q$-vector phonon by the splitting (confluence) processes. The difference of the transition probabilities can be deduced as
\begin{eqnarray}
p^{(s)}_{3}(k+q, q)-p^{(c)}_{3}(k, q)
\approx-p^{(c)}_{3}(k,q)\frac{\hbar \omega^{(p)}_{q}}{k_{B}T_{p}}\frac{\Delta T_{pm}}{T_{p}},
\end{eqnarray}
we have then}
\begin{eqnarray}
\tau_{mp}&\approx&\frac{c_{p}c_{m}}{c_{T}}\frac{k_{B}T^{2}}{\sum_{k,q}p^{(c)}_{3}(k,q)[\hbar \omega^{(p)}_{q}]^{2}}.
\label{eq:ralaxation-time-of-phonon-magnon-system-1}
\end{eqnarray}
By using Eq.(\ref{eq:jumping-probabilities-of-three-particle-1}) and Eq.(\ref{eq:ralaxation-time-of-phonon-magnon-system-1}), the thermal relaxation time have been numerically computed. As demonstrated in the  Fig.(\ref{fig:relaxation-time-for-three-particle-process}), at very low temperatures (1-10K), $\tau_{mp}(T)$ decrease very fast. However, $1/\tau_{mp}(T)$ becomes a linear function in the high temperature regime  (over 70K), which is different with the $T^4$-dependence in Ref.[\onlinecite{Kaganov1959}]. It may be due to the effect of dimensionality: one dimensional Cherenkov processes in our case but three-dimensional magnons/phonons in M. I. Kaganov and V. M. Tsukernik's studies.

\section{Temperature distribution of long wavelength magnons}
Given that the magnon-phonon coupling is wave number dependent, the thermal relaxation processes between the magnon and phonon baths are expected to be dominated by the short wavelength magnons,  as indicated by the experimental observations in the Ref.[\onlinecite{{Agrawal2013}}].  The temperature distribution of magnon subsystem should display a strong spectral dependence as the ferromagnetic chain  subject to a thermal (phonon) gradient.

Considering that a ferromagnetic chain is contacting to the left and right thermal sources with temperature $T_{L}$ and  $T_{R}$ at the position of $-\frac{L}{2}$ and $\frac{L}{2}$, respectively, the dynamic thermal equilibria can be established energetically and firstly between the phonons and the short wavelength magnons. The temperature distribution of the short wavelength magnon system is given by the SW model \cite{Sanders1977}:
\begin{eqnarray}
T_{m}(x)&=&T_{0}-\frac{Q}{K_{T}}[x-\frac{\sinh{Ax}}{A\cosh{\frac{1}{2}AL}}].
\label{eq:distribution-of-main-body-magnons}
\end{eqnarray}
where $T_{0}=(T_{L}+T_{R})/2$ is the temperature of middle point $x=0$, $L$ is the length of the sample, and $Q$ is the total heat current. $K_{T}=K_{p}+K_{m}$ is the total thermal conductivity, $K_{p}$ and $K_{m}$ are the thermal conductivity of phonons and magnons, respectively.
\begin{eqnarray}
A^{2}&=&\frac{1}{\lambda^{2}}=\frac{C_{m}C_{p}}{C_{T}}\frac{K_{T}}{K_{m}K_{p}}\frac{1}{\tau_{mp}}
\label{eq:A-lambda}
\end{eqnarray}
is a factor related to relaxation times with the dimension $[L]^{-2}$, $C_{m}, C_{p}, C_{T}$ are the heat capacities of the magnons, phonons and the magnon-phonon system on unit length. With a strong coupling strength between the phonons and short wavelength magnons and thus small $\tau_{mp}$, $T_{m}(x)$ becomes very close to the temperature profile of phonons.

As for the magnonic relaxation processes, the rate of the energy gain of (relatively) short wavelength $k$-magnons in the region of $[x,x+dx]$ is
$\frac{d\mathcal{Q}_{k}(x)}{dt}dx\approx\frac{c_{mk}}{\tau_{mk}}[T_{m}(x)-T_{mk}(x)]\frac{dx}{L}
=\frac{C_{mk}}{\tau_{mk}}[T_{m}(x)-T_{mk}(x)]dx$, where $\mathcal{Q}_{k}(x)$ is the energy gain of $k$-magnons on unit length, $c_{mk}$ is the total heat capacity of $k$-magnons, and $C_{mk}=c_{mk}/L$ is the heat capacity on unit length. $T_{m}(x)$ and $T_{mk}(x)$ here are not the temperatures of the whole system, but defined as the local model that have been proposed by Xiao \emph{et al}\cite{XiaoJiang2010}. Then the rate of the energy gain of $k$-magnons in the region  $[-\frac{L}{2},x]$ can be derived as
\begin{eqnarray}
Q_{k}(x)&=&\frac{C_{mk}}{\tau_{mk}}\int^{x}_{-L/2}[T_{m}(x')-T_{mk}(x')]dx',
\label{eq:heat-flux-of-k-vector-magnons}
\end{eqnarray}
which should be equivalent to the energy current of the $k$-magnons at position $x$, {$Q_{k}(x)=-K_{mk}dT_{mk}(x)/dx$}, where $K_{mk}$ is the thermal conductivity of $k$-magnons. Differential equation of temperature distribution for $k$ magnons is obtained,
\begin{eqnarray}
\frac{d^{2}T_{mk}(x)}{dx^{2}}+\frac{C_{mk}}{K_{mk}\tau_{mk}}[T_{m}(x)-T_{mk}(x)]=0.
\label{eq:equation-of-temprature-distribution-of-k-magnons}
\end{eqnarray}
Taking the fact that magnons can not exchange energy with thermal sources directly \cite{Sanders1977}, one has the boundary conditions $dT_{mk}(-L/2)/dx = dT_{mk}(L/2)/dx =0$ and $T_{mk}(0)=T_{0}$. Solving  Eq.(\ref{eq:equation-of-temprature-distribution-of-k-magnons}) with Eq.(\ref{eq:distribution-of-main-body-magnons}), it gives
\begin{eqnarray}
T_{mk}(x)=T_{0}-\frac{Q}{K_{T}}
\left[x-\frac{(1-\beta)\sinh{Bx}}{B\cosh{\frac{1}{2}BL}}
-\frac{\beta\sinh{Ax}}{A\cosh{\frac{1}{2}AL}}\right]
\label{eq:equation-of-temprature-distribution-of-magnons}
\end{eqnarray}
with
\begin{eqnarray}
B^{2}&=& \frac{C_{mk}}{K_{mk}}\frac{1}{\tau_{mk}} \hspace{0.4cm} \text{and} \hspace{0.4cm}
\beta=-\frac{B^{2}}{A^{2}-B^{2}}\hspace{0.3cm} (A\neq B).
\end{eqnarray}
When $A =B$, then
\begin{eqnarray}
&&T_{mk}\left( x\right)  =T_{0}-\frac{Q}{K_{T}}\times\\
&&\left[x-\left( \frac{3}{2}+\frac{1}{2}\frac{AL/2\sinh AL/2}{\cosh AL/2}\right)
\frac{\sinh Ax}{A\cosh AL/2}
+\frac{1}{2}\frac{x\cosh Ax}{\cosh AL/2}\right].\nonumber
\label{eq:equation-of-temprature-distribution-of-magnons-1}
\end{eqnarray}
\begin{figure}[b]\centering
\includegraphics[scale=0.33]{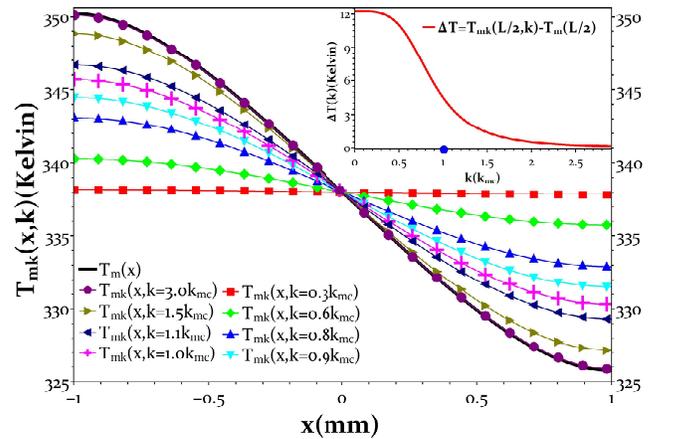}
\caption{Temperature distributions of $k$-magnons with $\lambda=0.47$ mm and $L=2$ mm. 
Inset shows the temperature deviations at $x=L/2$, where the characteristic wave vector  $k_{mc}$ is given by $B=A$.}
\label{fig:temperature-of-longwave-magnons}
\end{figure}

We adopt the one-dimensional thermal conductional formula\cite{C. Kittel, Huangkun1988} $K_{mk}=C_{mk}v_{mk}l'_{mk}$, where $l'_{mk}$ is the mean free path of $k$-magnons, {and} $v_{mk}$ is the velocity of $k$-magnons. For long wavelength magnons, $v_{mk}\approx \frac{\mathcal{J}}{2}a^{2}k/\hbar, l'_{mk}=v_{mk}\tau'_{mk}=\frac{\mathcal{J}}{2}a^{2}k\tau'_{mk_{0}}n^{(m)}_{k}/(\hbar n^{(m)}_{k_{0}})$, where $k_{0}$ is the minimal wave vector of magnons. We have then
{\begin{eqnarray}
B^{2}&\approx&\frac{\hbar^{2}}{\tau'^{2}_{mk_{0}}[\hbar \omega^{(m)}_{k_{0}}]^{2}}
\frac{1}{n^{(m)}_{k}}k^{2}\approx \left[\frac{\hbar/\tau'_{mk_{0}}}{\hbar \omega^{(m)}_{k_{0}}} \right ]^{2} \frac{\mathcal{J } a^{2}}{2k_{B}T_{m}}k^{4}.
\end{eqnarray}}

The $k^{-4}$-dependence of relaxation time $\tau_{mk}$ (cf. Eq.\ref{eq:ralaxation-of-k-wave-vector-magnons}) implies that $\tau_{mk}$ reduces rapidly with increasing $k$. For the $k$-magnons with relatively short wavelength,  $\tau_{mk}$ can be much smaller than $\tau_{mp}$, then $B\gg A$ and $\beta\sim 1$, Eq.(\ref{eq:equation-of-temprature-distribution-of-magnons}) returns to $T_{mk}(x) \approx T_{m}(x)= T_{0}-\frac{Q}{K_{T}}[x-\frac{\sinh{Ax}}{A\cosh{\frac{1}{2}AL}}]$. Whereas, when the relaxation time {of the $k$-magnons} is much larger than $\tau_{mp}$, one has then $A\gg B$ and $\beta=-B^{2}/(A^{2}-B^{2})\approx 0$, Eq.(\ref{eq:equation-of-temprature-distribution-of-magnons}) becomes
$T_{mk}(x)\approx T_{0}-\frac{Q}{K_{T}}[x-\frac{\sinh{Bx}}{B\cosh{\frac{1}{2}BL}}]$. Furthermore, as $BL\sim 0$, we have $T_{mk}(x)\sim T_{0}$.
The detailed characterization of temperature distribution is worked out numerically for varying wave number $k$ of magnons in
Fig.(\ref{fig:temperature-of-longwave-magnons}). It shows evidently that the temperature profile of $k$-magnons is largely lifted up in the cooler regions and deeply pulled down in the hotter ones as the wave number $k$ of magnon decreasing. There is a  characteristic wave vector $k_{mc}$ that is determined by the condition $B^{2}= A^{2}$ as,
\begin{eqnarray}
k_{mc}&=&\sqrt[4]{\left[\frac{\hbar \omega^{(m)}_{k_{0}}}{\hbar/\tau'_{mk_{0}}}\right]^{2}\frac{2k_{B}T_{m}}{\mathcal{J}a^{2}}A^{2}}.
\label{eq:critial-wave-vector-of-magnons}
\end{eqnarray}
Below $k_{mc}$, the temperature deviations of $k$-magnons becomes more pronounced. Such strong temperature inequality of long wavelength magnons could be responsible for the spin Seebeck effect once the spin pumping is proportional to the  difference of the temperatures between magnons and phonons\cite{XiaoJiang2010}.

From the experimental point of view, it is important to estimate the characteristic wave vector $k_{mc}$ in yttrium iron garnet (YIG) that is an excellent candidate to study the  spin-Seebeck effect. Taking $k_{0}\sim 10^{5}$ rad/m around room temperature, together with $a=1.24$ nm{\cite{Ruckriegel2014}}, {$\mathcal{J}\sim 19 $ meV}\cite{Ruckriegel2014,Diniz2016}, $T_{m}=300$ K, and $A^{2}=4.5 \times 10^{6} $m$^{-2}$ ($\lambda=1/A=0.47$mm\cite{Agrawal2013}), the characteristic wave vector is estimated as {$k_{mc}\approx 1.68\times 10^{8}~ \text{rad/m}$}, which is in good agreement with the experimental observations.

\section{Conclusion}
We have presented that the magnon-phonon coupling, the magnon transition probability, the magnon relaxation processes, and the magnon temperature profile are all spectral dependent. There is a characteristic magnon wave vector $k_{mc}$ that is determined by the competition of magnon-phonon and magnon-magnon relaxation processes. For the magnons with wave-vector $k < k_{mc}$, the inequality between the temperatures of $k$-magnon and the phonon bath becomes increasingly pronounced with lower magnon wave-vector $k$.  As proposed by M. Agrawal  \emph{et al}\cite{Agrawal2013} and other research groups, such considerable magnon/phonon temperature difference should be of particular importance to the spin Seebeck effect.

\section{ACKNOWLEDGEMENTS}
This work is supported by the National Natural Science Foundation of China (No. 11474138), the German Research Foundation (No. SFB 762), and the Program for Changjiang Scholars and Innovative Research Team in University (No. IRT-16R35), and the  Science Foundation of North Minzu university (Grant No. 2017DX006, China).

\end{document}